\documentclass[journal]{IEEEtran}

\usepackage{amsmath}
\usepackage{graphicx}
\usepackage{algorithm}
\usepackage{algorithmic}
\usepackage{subcaption}

\begin{document}

\title{Coordination of DERs for Grid Reliability via Day-ahead Demand-Supply Power Bounds}

\author{Thomas Navidi, 
Abbas El Gamal, \textit{Life Fellow, IEEE}, and 
Ram Rajagopal, \textit{Member, IEEE}

\vspace{-20pt}
\thanks{T. Navidi and A. El Gamal are with the Department of Electrical Engineering at Stanford University. 
R. Rajagopal is with the Departments of Civil and Environmental Engineering and Electrical Engineering at Stanford University.}
}

\maketitle

\begin{abstract}
A previous study has shown that coordinating DERs to protect the distribution grid can significantly reduce the infrastructure upgrades needed to address future increases in DER and electrification penetrations. Implementing such coordination in the real world, however, is challenging due the temporal and spatial uncertainties about the loads and renewable generation, smart meter and network delays, incomplete information about the grid, different consumer objectives and privacy constraints, and scalability of the coordination scheme. This paper describes a day-ahead 2-layer DER coordination scheme that addresses these challenges. A global controller uses historical load data to compute day-ahead hourly demand upper and lower bounds for each consumer node. It then solves a largest volume axis-aligned box optimization problem to determine corresponding supply power bounds which if followed, ensures grid reliability. A local controller at each consumer node then determines the DER power injections which satisfy the consumer's objectives while obeying its supply bounds. Simulation results demonstrate, for example, that this scheme can capture 62\% of the reduction in transformer violations achievable by the perfect-foresight centralized controller used in the aforementioned previous study.
\end{abstract}


\IEEEpeerreviewmaketitle

\section{Introduction}

Coordination of the power injections from distributed energy resources (DER), such as EV chargers, battery storage units, and flexible thermal loads, across a distribution network can provide several benefits, including reducing peak network load during extreme weather conditions, providing grid services, and safeguarding grid infrastructure, such as transformers and voltage regulators. Existing programs for DER coordination, such as virtual power plants (VPPs) and demand response, aim to provide grid services but do not consider distribution grid reliability. In a recent study~\cite{joule}, the authors showed that coordinating DERs for grid reliability has the potential to significantly reduce the grid infrastructure upgrades needed to support future increases in DER and electrification penetrations, while also reducing peak network load. The perfect foresight centralized controller used in~\cite{joule} to demonstrate these benefits, however, is not implementable and it has remained unclear how much of these benefits can be attained by practical coordination schemes which must operate under challenging requirements, including: (i) temporal and spatial uncertainties about the loads and renewable generation, (ii) smart meter and communication delays that can be up to several hours~\cite{smart_meter_issues} as well as communication network delays, (iii) incomplete information about the physical characteristics of the network, (iv) different objectives and privacy constraints of the consumers who own the DERs, and (v) scalability to large numbers of consumers.

In this paper, we present a day-ahead DER coordination scheme whose main objective is to minimize transformer and voltage violations across the distribution network under the aforementioned real-world constraints. We demonstrate that this scheme can reap a significant fraction of the benefits of the perfect foresight controller with a moderate increase in consumer electricity cost. To address the above first two requirements of an implementable scheme, we use the 2-layer control architecture first reported in~\cite{kyle_anderson:2017} which comprises a global controller (GC) that could be operated by the DSO or via an aggregator cooperating with the DSO, and local controllers (LC) located at consumers' sites. Communication takes place only between the GC and the LCs and the GC sends updates to the LCs once per a day. Each day, the GC uses past power injection data obtained from the smart meters to compute hourly power \emph{demand} upper and lower bounds over the following day for every consumer or group of consumers in the network. The GC then computes corresponding hourly power \emph{supply} upper and lower bounds by solving a maximum volume axis aligned box optimization problem~\cite{largest_inscribed_rect} whose objective is to maximize the volume of the box defined by the supply bounds subject to power flow constraints and fairness across consumers while staying within the given upper and lower power demand bounds. This formulation guarantees that any set of power injections that satisfies the supply bounds does not violate the reliability constraints. The supply power bounds for each consumer node is then sent to its LC to determine the power injections over the span of the following day by minimizing a combination of its own objective and deviation from the supply bounds subject to its DER constraints. The scheme satisfies the third requirement above by using a data-driven linear model of the network learned through smart meter data. It further satisfies the fourth requirement by not requiring knowledge of any consumer constraints, objectives or DER data, such as SOC of storage, flexible load, or EV charging rate. The scheme is scalable, satisfying the last requirement for implementable coordination, since the maximum volume axis aligned box problem is convex  and the constraint space is independent of the number of DERs in the network.

This paper provides several significant enhancements and improvements over our earlier conference paper~\cite{Navidi_smartgridcomm}. While in~\cite{Navidi_smartgridcomm} the supply bounds were computed using a heuristic, the new GC algorithm provides guarantees on grid reliability when the bounds are satisfied and the linear power flow model is accurate. This paper also considers flexible thermal load as another controllable DER in addition to the EV charging and storage considered in~\cite{Navidi_smartgridcomm}
While~\cite{Navidi_smartgridcomm} included very limited simulation results, this paper provides extensive results for 11 3-phase distribution networks representing varying climates, sizes, and mixtures of commercial and residential consumers over a 3-decade horizon with recent projections of DER penetrations from~\cite{NREL_EV_adoption, NREL_EV, NREL_solar, NREL_storage}. We further benchmark the performance of our scheme relative to the bookend local and perfect foresight centralized  controllers in~\cite{joule}.

In addition to~\cite{kyle_anderson:2017,Navidi_smartgridcomm}, there has been other DER coordination schemes, sometimes referred to as management systems (DERMS), that aim to protect the grid. For example~\cite{price_DLMP} describes a hierarchical control scheme in which a centralized controller uses power flow to calculate distribution grid locational marginal prices (DLMPs) for each DER. Each local DER controller then attempts to minimize its assigned DLMP. This scheme, however, assumes that the objective of all consumers is to minimize the DLMP rather than allowing each of them to define its own objective. The strategy in~\cite{baosen_inverter_coord} uses a centralized controller to determine Lipschitz constraints that guarantee the safe operation of a local reinforcement learning algorithm for voltage stability. Their control strategy, however, is only applied to voltage stability and not to other objectives such as transformer overloading or individual DER objectives and is only applicable to a neural network based local controller. In~\cite{flex_provision} the authors present a method to calculate flexibility bounds at a single point in the distribution grid based on voltage limits. They do not consider transformer limits and do not fully explore the extension of their method to calculating bounds for multiple points in the grid simultaneously. Finally, the authors in~\cite{NREL_bounds} determine bounds for each node by solving a similar maximum volume box problem. In contrast to the work presented in this paper, however, their paper (i) only considers voltage constraints, (ii) the bounds are determined based only on the network model with no consideration of consumer loads (e.g., demand bounds), which can lead to unfairly allocating too much flexibility to nodes with smaller loads at the expense of nodes with larger loads, (iii) does not include any local controllers, (iv) has very limited simulation results, and (v) does not provide any performance evaluation relative to other DER coordination schemes.  

In the following section we introduce the grid and DER models and the various assumptions we make in formulating the bounds coordination problem. In Section~\ref{sec:scheme}, we describe the bounds scheme. In Section~\ref{sec:setup}, we present the simulation setup used in evaluating our scheme. Since this setup is the same as that in~\cite{joule}, we provide only a brief summary and refer the readers to that paper for more details. In Section~\ref{sec:results}, we compare the performance of our scheme to the local and the perfect-foresight controllers used in~\cite{joule} to bookend the performance of any implementable DER coordination scheme. 

\section{Models and assumptions}
\label{sec:formulation}

We consider control of distributed energy resources within a 3-phase distribution network modeled by a graph with a set of nodes $[1:N]$  and a set of edges (lines) $\mathcal{L}$. Node $i=1$ corresponds to the substation which is connected to the transmission system and supplies power to the rest of the nodes. We denote the set of consumer nodes by $\mathcal{C} \subset [1:N]$, where $|\mathcal{C}| = N_\mathrm{C}$. The set of transformers in the network is specified by a set of node pairs $(i,j) \in \mathcal{T}$, where and $|\mathcal{T}|=N_\mathrm{T}$. Each consumer node comprises a collection of uncontrollable loads, solar PV, energy storage, EV chargers, and flexible thermal loads located under a single secondary transformer or a single smart meter. This collection can represent one or several residential consumers or a larger commercial building. We consider the steady state operation of the network over time steps $t\in \{1,2,\ldots\}$ each of length $\delta$, which we assume to be $15$min in our simulations. The total complex net power consumption at each node $i\in [1:N]$ in the network in timestep $t$ is denoted by $s_{i}(t)$. The steady state power flows in the network determine the voltage magnitude $v_{i}(t)$ for each node $i$ at time $t$.

In the following we provide the models and assumptions on the loads and controllable DERs we assume throughout the paper. 
 
\noindent{\bfseries Uncontrollable load}. This includes a stochastic uncontrollable load, with constant power factor, and solar PV generation, which is assumed to be real and independent of the power factor. The combined uncontrollable load at node $i$ and timestep $t$ is denoted by $p_i(t)$.

\noindent{\bfseries Energy storage}. We assume the battery storage $k$ at node $i$ has maximum capacity (energy) $Q_{ik}^{\mathrm{max}}$ At time $t$, the storage unit at node $i$ has a charging power rate of $c_{ik}(t)\ge 0$, discharging rate of $d_{ik}(t)\ge 0$, and a state of charge (energy) of $Q_{ik}(t)$. We assume that each storage unit only charges and discharges real power, and that storage charges and discharges with efficiency $0 \le \gamma^c_{ik} \le 1$ and $0 \le \gamma^d_{ik} \le 1$, respectively. Hence, the storage energy at time $t$ is modeled by $Q_{ik}(t) = Q_{ik}(t-1) + \gamma^c_{ik} \delta\, c_{ik}(t) - \gamma^d_{ik} \delta\, d_{ik}(t)$. The maximum charging and discharging energy rates are denoted by $c_{ik}^{\mathrm{max}}$ and $d_{ik}^{\mathrm{max}}$, respectively.

\noindent{\bfseries Electric vehicle charging}. The model we use for EV charging is similar to that for energy storage with maximum discharge power $d_{ik}^{\mathrm{max}}=0$ (we do not consider vehicle to grid in this paper) and $c_{ik}(t)$ is bounded above by the EV charger rated power. Each EV $k \in \mathcal{E}_i$, where $\mathcal{E}_i$ is the set of EVs at node $i$, is to be charged within non-overlapping time windows $\mathcal{W}_{ik}$ over the simulation time horizon. The end time of window $w \in \mathcal{W}_{ik}$ is denoted by $t^{\mathrm{end}}_w$. An additional constraint on EV charging comes from the requirement that it must be charged to a desired level $Q_{ikw}^{\mathrm{final}}$ by the end of each window, i.e., $Q_{ik}(t^{\mathrm{end}}_w) = Q_{ikw}^{\mathrm{final}}$, $w \in \mathcal{W}_{ik}$ and $k \in \mathcal{E}_i$.

\noindent{\bfseries Flexible thermal load.} These include loads for air conditioning, ventilation, electric space and water heating and cooking. We denote the total thermal load (power consumption) at node $i$ and timestep $t$ by $u_{i}(t)$ and its maximum over the simulation horizon by $u_i^{\mathrm{max}}$, and assume that it has a constant power factor of $0.92$. We further denote the fraction of flexible thermal energy that can be shifted to another time within the same day by $\phi$, which depends on the network and whether the node is commercial or residential. We express the thermal energy at time step $t \in [1:T_d]$ as $Q_{ik}(t) = Q_{ik}(t-1) + \delta\, c_{ik}(t)$, where $Q_{ik}(0)=0$. In practice, the thermal energy consumed by the flexible load is determined by the consumers desired temperature band for each thermal load. As the appliance consumes energy, the temperature either increases or decreases. To provide grid services, the controller adjusts the thermal load energy consumption while keeping the temperature within the band. Accurately relating the thermal energy consumption to temperature can be very complex, requiring detailed data about the appliances, thermal models of the buildings, climate conditions, and consumer behavior. Since this paper is focused on the bounds coordination scheme and on comparing its performance to the bookend controllers in~\cite{joule} rather than on thermal load modeling, operation, and consumer behavior, we will use the simplified temperature to thermal load energy consumption model in Section~\ref{sec:lc}. 

In summary, the dynamics for the battery storage, EV charging, and flexible load can all be expressed as $Q_{ik}(t) = Q_{ik}(t-1) +  \gamma^c_{ik} \delta\, c_{ik}(t) - \gamma^d_{ik} \delta\, d_{ik}(t)$, where $d_{ik}(t) = 0$ for EV charging and $\gamma^c_{ik} = 1$ and $d_{ik}(t) = 0$ for flexible thermal load with the additional constraints for each DER as described above.

\subsection{Linearized inverse power flow}
To maintain reliable operation, the network operator needs to manage both the voltage magnitudes at each node and the complex power flowing through each transformer to keep them within their respective safe ranges. The relationship between the net load and the voltage at each node $i\in [1:N]$ is governed by the AC power flow equations and accompanying constraints~\cite{low_exact_relaxation}. We denote the maximum rated apparent power squared of transformer $(i,j)$ by ${\tau}_{ij}^{\mathrm{max}}$, and the upper and lower limits on The voltage magnitude $v_i$ at node $i$ by $v_i^{\mathrm{min}}$ and $ v_i^{\mathrm{max}}$. We say that a set of power injections $s_i$ for $i\in [1:N]$ is feasible if it satisfies these transformer and voltage limits.

In general, the power flow equations are non-convex which can lead to computationally intractable optimization problems. Under some conditions, this can be addressed through convex relaxation; however, the commonly used SDP and SOCP relaxations do not provide a correct solution when the objective function is a decreasing function of the real power injection as is the case for the maximum volume box problem discussed later in this paper~\cite{low_exact_relaxation}. This may be alleviated through the use of sequential convexification and modifications to the relaxation via the method described in~\cite{sequential_convex}. However, this method still requires knowledge of the physical parameters of the network, which are often not available even to the DSO. Moreover, since we are only interested in scheduling, as in other settings such as day ahead markets, we use the following linearized the power flow equations~\cite{LinearPF_LSQ, LinearPF_Jiafan} 
\begin{subequations}
\begin{align}
    \boldsymbol{v}(t) &= A \boldsymbol{s}(t) + \boldsymbol{a}, \\
	\boldsymbol{\tau}(t) &= (F\boldsymbol{s}(t) + \boldsymbol{f})^2 + (G \boldsymbol{s}(t) + \boldsymbol{g})^2,
 \end{align}
\end{subequations}
where the bold symbols refer to the vector versions of the variables defined earlier, $A$, $\boldsymbol{a}$, $F$, $\boldsymbol{f}$, $G$, $\boldsymbol{g}$ are the linear model coefficients learned from the data, and the square of a vector refers to an element-wise square operation.

In addition to significantly reducing the computational complexity of computing the supply bounds, this model can in fact be more accurate than physics based approximations~\cite{LinearPF_LSQ} because the parameter values of such models, such power line admittances, configurations and parameters of capacitor banks, voltage regulators, and switches, are often not completely known to the GC whose function is to characterize the set of feasible power injections via the supply bounds. In contrast, the power and voltage measurement data needed to train the above model are readily available through existing smart metering infrastructure. Furthermore, this data driven model can readily adapt to changes in the states of the network assets, such as voltage regulators and switches, without needing to know the specifics of such changes.

\subsection{Reactive power modeling}
\label{sec:reactive_power}
Since the voltages, hence, the feasible set of real power injections, depend on the reactive power injections, to establish the real power injection supply bounds we need to know the relationship between the reactive power at each consumer node and its real power. Since the GC does not have information about the composition of the load at each node, we estimate this relationship using the historical net load data from the smart meters and the transformer monitoring equipment. Since the supply bounds need to guarantee that the complex power is within a certain feasible region, we use this relationship to establish an upper and lower percentile estimate of the reactive power as a function of the real power. The specific percentile can be adjusted to tune how conservative of an estimate to use for the range of possible reactive power injections.
In this study, we use a linear least squares estimator trained on the 90th percentile of reactive power injections for the upper bound and the 10th percentile for the lower bound. The resulting upper and lower bounds on the reactive power (as a function of the real power) are
\begin{subequations}
\begin{align}
    \boldsymbol{q}^\mathrm{u}(t) &= H^\mathrm{u}(t)\boldsymbol{p}^\mathrm{u}(t) + \boldsymbol{h}^\mathrm{u}(t),\\ 
    \boldsymbol{q}^\mathrm{l}(t) &=  H^\mathrm{l}(t)\boldsymbol{p}^\mathrm{l}(t) + \boldsymbol{h}^\mathrm{l}(t),
    \label{eq:reactive_model}
\end{align}
\end{subequations}
where $\boldsymbol{q}^\mathrm{u}(t)$ and $\boldsymbol{q}^\mathrm{l}(t)$ are the upper and lower bounds on reactive power, and $H^\mathrm{u}, \boldsymbol{h}^\mathrm{u},H^\mathrm{l}, \boldsymbol{h}^\mathrm{l}$ are the linear model coefficients. The model coefficients are a function of $t$ since there are different coefficients for each hour in the day.

\section{Day-ahead DER coordination scheme} 
\label{sec:scheme}

\begin{figure}[htpb]
	\centering
	\includegraphics[scale=0.68]{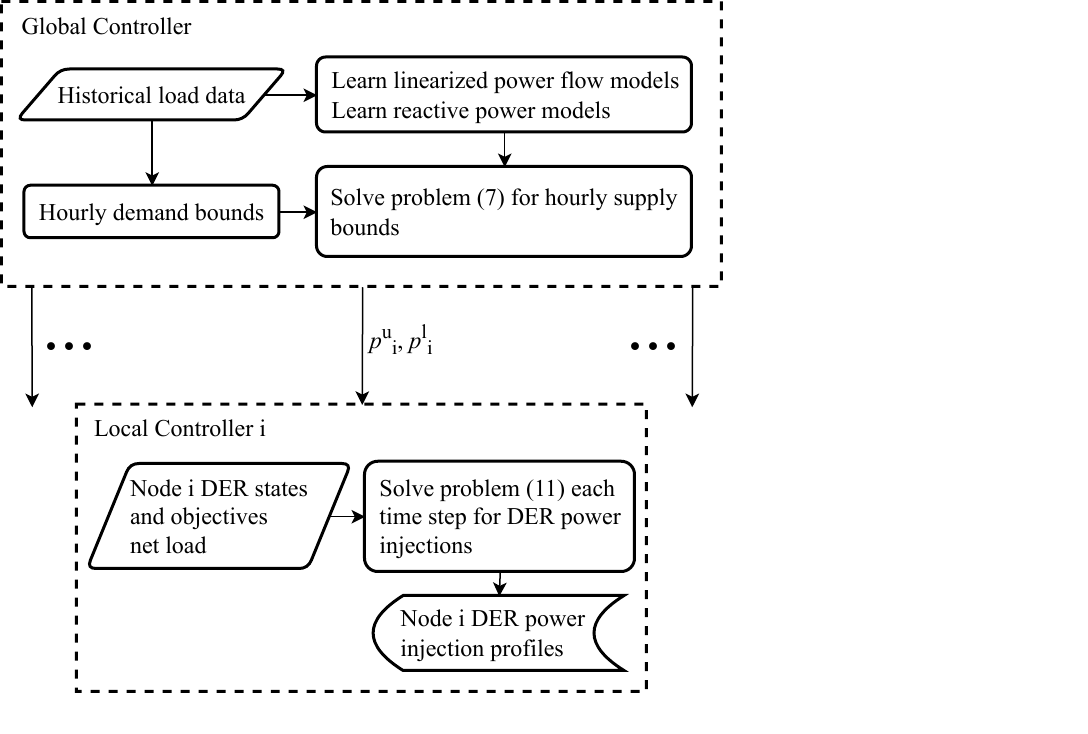}
	\caption{Block diagram of the day ahead two-layer coordination scheme. Input data are described in the parallelograms and operations in the rectangles.}
	\label{fig:block_diagram_bounds}
\end{figure}

A block diagram of the day ahead 2-layer DER coordination scheme is given Figure~\ref{fig:block_diagram_bounds}. In the following subsections we describe the GC and the LC in detail.

\subsection{Global controller}
The AC power flow equations provide a mapping from the nodes real and reactive power injections to their voltages, hence the set of feasible node voltages, i.e., voltages within tolerable limits, can be represented by a set of feasible power injections. Similarly, the transformer constraints represent another set of feasible power injections. The intersection of these two sets represents the feasible set of power injections in the network, that is, the power injections that can be reliably supported by the distribution network. Since each consumer's LC has access only to its own power injections and does not have any information about the power injections at other nodes, the LCs cannot jointly operate at any arbitrary point in this feasible set of power injections. Hence in our scheme, the GC provides each LC with day-ahead hourly ``supply" upper and lower bounds denoted by the $24$-dimensional vectors $\boldsymbol{p}_i^\mathrm{l}$ and $\boldsymbol{p}_i^\mathrm{u}$ on its net power injections for each node $i \in \mathcal{C}$ such that if every LC operates within its given supply bounds, the resulting set of power injections in the network is feasible. Such bounds define a ``supply box" in the $N_\mathrm{C}$-dimensional space of consumer node power injections for each hour of the next day. 

\begin{figure}[htpb]
	\centering
	\includegraphics[scale=0.37]{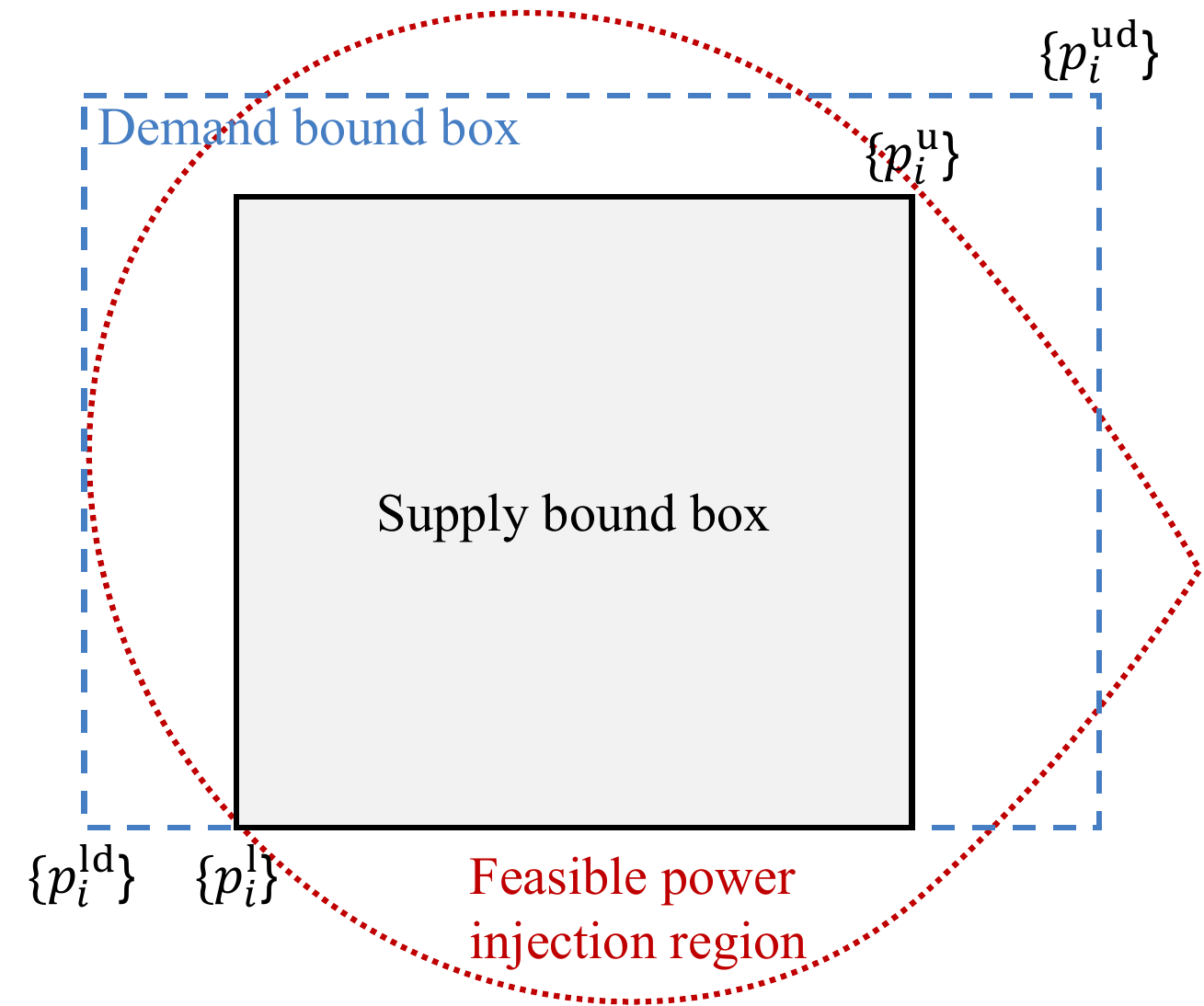}
	\caption{Illustration of the maximum volume axis aligned box problem.}
	\label{fig:bounds_space}
\end{figure}

Determining the supply bounds by finding the axis-aligned supply box with the maximum volume that fits within the set of feasible power injections, however, can lead to situations in which consumers in favorable locations in the network receive significantly more power injection flexibility relative to their actual demands, resulting in much higher inevitable supply bound violations at other nodes. To address this problem, node demands must be taken into consideration in finding the supply bounds. This is achieved by first finding upper and lower hourly ``demand" bounds $\boldsymbol{p}^\mathrm{ld}_i, \boldsymbol{p}^\mathrm{ud}_i$ on the net power injections for each node $i \in \mathcal{C}$ and for each hour of the next day as follows. To find the hourly demand bounds for a consumer node, the GC uses the node's historical load data corresponding to the next day, for example, if the next day is a weekday, the load data would be from the past weekdays in the past 5 weeks, similarly for weekends. The GC then sets the lower demand bound for each hour to be the lowest of either 0 or the power injection value of the selected historical data for that hour, and the upper demand bound for each hour is the highest of either 1kW or the power injection value of the selected historical data for that hour. The derived demand bounds define a demand box in the $N_\mathrm{C}$-dimensional space of consumer node power injections for each hour of the following day; see Figure~\ref{fig:bounds_space}.

From the above discussion, to ensure maximum flexibility the GC aims to find the largest volume axis-aligned supply box (defined by the supply bounds) which is in the intersection of the axis-aligned demand box and the set of feasible power injections. Further to prevent unfairly allocating more flexibility to nodes with lower demands at the expense of other nodes, we normalize the difference between the supply and demand bounds for each node by the difference between the upper and lower demand bounds for that node (for each hour) as follows
\begin{align}
    \Delta_i^\mathrm{u} = \frac{p_i^{\mathrm{ud}} - p_i^{\mathrm{u}}} {p_i^{\mathrm{ud}} - p_i^{\mathrm{ld}}},\; 
    \Delta_i^\mathrm{l} = \frac{p_i^{\mathrm{l}} - p_i^{\mathrm{ld}}} {p_i^{\mathrm{ud}} - p_i^{\mathrm{ld}}},\; i \in \mathcal{C},
\end{align}
where $0 \le \Delta_i^\mathrm{u},\Delta_i^\mathrm{l}<1$ and $\Delta_i^\mathrm{u} + \Delta_i^\mathrm{l} < 1$. The reason this normalization ensures fairness in flexibility across all consumer nodes is that the maximum box volume solution tries to make the supply box as close as possible to a hypercube within the demand box, such that shorter demand box dimensions have their corresponding supply box side lengths prioritized over the longer dimensions. This leads to the following supply bounds optimization problem formulation for each hour of the following day (we do not introduce an index for hour).{\allowdisplaybreaks
\begin{subequations} \label{eq:bounds_opt}
	\begin{alignat}{2}
		\text{maximize:} \;\; & \sum_{i\in \mathcal{C}} \log (1 - \Delta_i^\mathrm{u} - \Delta_i^\mathrm{l}) \label{eq:obj}\\
		\text{subject to:} \;\; & \Delta_i^\mathrm{u} \geq 0,\; \Delta_i^\mathrm{l} \geq 0,\; \Delta_i^\mathrm{u} + \Delta_i^\mathrm{l} < 1, \; i \in \mathcal{C}  \label{eq:bounds_1} \\
		& p_i^{\mathrm{u}} = p_i^{\mathrm{ud}} - (p_i^{\mathrm{ud}} - p_i^{\mathrm{ld}})\Delta_i^\mathrm{u},\nonumber \\
		& p_i^{\mathrm{l}} = p_i^{\mathrm{ld}} + (p_i^{\mathrm{ud}} - p_i^{\mathrm{ld}})\Delta_i^\mathrm{l},\; i \in \mathcal{C} \label{eq:bounds_4} \\
		& \boldsymbol{q}^{\mathrm{u}} = H^\mathrm{u}\boldsymbol{p}^\mathrm{u} + \boldsymbol{h}^\mathrm{u},\;
         \boldsymbol{q}^{\mathrm{l}} = H^\mathrm{l}\boldsymbol{p}^\mathrm{l} + \boldsymbol{h}^\mathrm{l}, \label{eq:reactive_1} \\
        & \boldsymbol{v} = A \boldsymbol{s} + \boldsymbol{a},\; \boldsymbol{s} \in [\boldsymbol{s}^{\mathrm{l}}, \boldsymbol{s}^{\mathrm{u}}] \label{eq:false_1}\\
        & v_i^{\mathrm{min}} \leq v_{i} \leq v_i^{\mathrm{max}},\; i \in [1:N] \\
        & \boldsymbol{\tau} = (F\boldsymbol{s} + \boldsymbol{f})^2 + (G \boldsymbol{s} + \boldsymbol{g})^2, \; \boldsymbol{s} \in [\boldsymbol{s}^{\mathrm{l}}, \boldsymbol{s}^{\mathrm{u}}] \\
        & \tau_{ij} \leq \tau^{\mathrm{max}}_{ij},\; (i,j) \in \mathcal{T}. \label{eq:false_8}
	\end{alignat}
\end{subequations} }
The objective function~\eqref{eq:obj} is the sum of the log of the normalized width for each node's supply bounds. Maximizing this objective is equivalent to maximizing the normalized volume of the axis aligned supply box. The constraints~\eqref{eq:bounds_1}-\eqref{eq:bounds_4} define the relationship between the demand and the supply bounds. The constraints~\eqref{eq:reactive_1} are the mappings from real power supply bounds to the reactive power injection bounds for each node. The constraints~\eqref{eq:false_1}-\eqref{eq:false_8} are the data-driven mappings of real and reactive power to the voltage magnitude and limits at each node and the apparent power and limits at each transformer, respectively. 

Solving this problem requires ensuring that every complex power injection $\boldsymbol{s} \in \big[\boldsymbol{s}^{\mathrm{l}}, \boldsymbol{s}^{\mathrm{u}}\big]$ satisfies the voltage and transformer constraints, hence the problem appears intractable. However, since both the set of power injections arising from the voltage constraints and from power injections corresponding to the transformer constraints are convex, their intersection with the demand box is also convex. Thus, we can take advantage of the convexity to simplify the problem as in~\cite{boyd_vandenberghe_2004} and its expanded version in~\cite{largest_inscribed_rect} in which the constraints reduce to those needed to define the convex feasible set. To do so, we split the power flow mapping matrices and coefficients into positive and negative components $A_+, A_-,\boldsymbol{a}_+$ and $\boldsymbol{a}_-$ such that $A_+ + A_- = A,\; \boldsymbol{a}_+ + \boldsymbol{a}_- = \boldsymbol{a}$,
and similarly for
$F_+, F_- \boldsymbol{f}_+, \boldsymbol{f}_-$ and $G_+, G_-, \boldsymbol{g}_+, \boldsymbol{g}_-$. The transformer mapping will be referred to as $B_+, B_-, \boldsymbol{b}_+, \boldsymbol{b}_-$ to simplify the notation. Then we can verify that every power injection within the supply bounds satisfies the constraints by only checking the upper and lower vertices of the box as follows
\begin{subequations}
    \begin{align}
        & \boldsymbol{v}^{\mathrm{u}} = A_+ \boldsymbol{s}^{\mathrm{u}} + \boldsymbol{a}_+  + A_- \boldsymbol{s}^{\mathrm{l}} + \boldsymbol{a}_-  \label{eq:network_1}\\
        & \boldsymbol{v}^{\mathrm{l}} = A_+ \boldsymbol{s}^{\mathrm{l}} + \boldsymbol{a}_+  + A_- \boldsymbol{s}^{\mathrm{u}} + \boldsymbol{a}_- \\
        & v_i^{\mathrm{min}} \leq v_{i}^{\mathrm{u}} \leq v_i^{\mathrm{max}},\;
		 v_i^{\mathrm{min}} \leq v_{i}^{\mathrm{l}} \leq v_i^{\mathrm{max}},\; i \in [1:N] \\
        & \boldsymbol{\tau}^{\mathrm{u}} = B_+ \boldsymbol{s}^{\mathrm{u}} + \boldsymbol{b}_+  + B_- \boldsymbol{s}^{\mathrm{l}} + \boldsymbol{b}_- \\
		& \boldsymbol{\tau}^{\mathrm{l}} = B_+ \boldsymbol{s}^{\mathrm{l}} + \boldsymbol{b}_+  + B_- \boldsymbol{s}^{\mathrm{u}} + \boldsymbol{b}_- \\
        & \tau_{ij}^{\mathrm{u}} \leq \tau^{\mathrm{max}}_{ij},\;
		\tau_{ij}^{\mathrm{l}} \leq \tau_{ij}^{\mathrm{max}},\; (i,j) \in \mathcal{T}. \label{eq:network_8}
    \end{align}
\end{subequations}

Replacing the constraints~\eqref{eq:false_1}-\eqref{eq:false_8} with the above constraints, yields a computationally efficient and scalable optimization problem which the GC solves for each hour of the next day. The resulting supply upper and lower bounds for each consumer node are then sent to its local controller. As long as each local controller is able to restrict the power injections to within its supply bounds, the network constraints are satisfied provided the linear mappings from the power injection to node voltages and transformer powers are accurate. 
\smallskip

\noindent{\bf Relationship to maximum loadability}. The maximum volume axis aligned box problem is related to the problem of finding the maximum loadability of a transmission system~\cite{sequential_convex, voltage_stability} for voltage stability analysis. The maximum loadability problem aims to maximize the power injection at a single node in the transmission grid constrained by the power flow equations and voltage limits. Key differences between this previous work and our box problem are: (i) we find both a maximum and minimum range of power injections for all nodes in the network rather than a single injection for all the nodes, (ii) we consider the variability of load over time to limit the power injection of nodes in the distribution network depending on their demands ahead of time, and (iii) we include transformer power limits in the constraint set.

\subsection{Local controller} \label{sec:lc}
The local controller determines the power injections for the stationary storage, EV chargers, and flexible thermal load within each consumer's node. It aims to satisfy the consumer's objectives and preferences, such as minimizing the total electricity cost, charging EVs on time, and maintaining comfortable climate while minimizing the deviation of the net power injection from its supply bounds. Why would consumers consider the supply bounds at the expense of less flexibility in satisfying their objectives, however? 

To address this key question, we propose incentivizing consumers via a tiered rate plan similar to ones currently offered by some utilities such as PG\&E~\cite{tiered_rate}. Under our proposed incentive, consumers would pay a reduced Time of Use (TOU) rate when their net power injection is within the supply bounds and regular TOU rate when they violate the bounds. The exact discounted rate, however, only needs to be sufficient for consumers to prioritize following the bounds over other cost saving strategies, such as energy arbitrage. 

Although our scheme allows for consumers to have different objectives, to evaluate the performance of the scheme, in Section~\ref{sec:results} we assume that all local controllers have the same objectives. In particular, to determine the power injections for the DERs at each node, the local controller solves the following optimization problem for every time step $t$, i.e., every $\delta=15$ minutes of each day.
\begin{subequations} \label{Localopt_op}
	\begin{alignat}{3}
 	\text{minimize:} &\;\; \lambda_b L_i^{b}(t) + \sum_{k=1}^{K_i} \lambda_{ik} L_{ik}^f(t) \\
    	\text{subject to:} &\;\;  0 \le \boldsymbol{c}_i(t) \le \boldsymbol{c}_i^{\mathrm{max}}(t),\;
     0 \le \boldsymbol{d}_i(t) \le \boldsymbol{d}_i^{\mathrm{max}}(t) \label{dmax} \\
    &\;\; \boldsymbol{Q}_i^{\mathrm{min}}(t) \le \boldsymbol{Q}_i(t) \le \boldsymbol{Q}_i^{\mathrm{max}}(t) \label{qmax} \\
&\;\; Q_{ik}(t) = Q_{ik}(t-1) + \gamma^c_{ik} \delta c_{ik}(t) \nonumber \\
&\qquad\qquad\qquad - \gamma^d_{ik} \delta d_{ik}(t), \; k \in [1:K_i] \label{charge} \\
   &\;\;  Q_{ik}(t^{\mathrm{end}}_w) = Q_{ikw}^{\mathrm{final}}, \; w \in \mathcal{W}_{ik}, k \in \mathcal{E}_i. \label{ev_charge}
	\end{alignat}
\end{subequations}
The first term in the above objective function corresponds to the deviation from the supply bounds:
\begin{multline}
 L_i^{b}(t) = \Big[ p_i(t) + \sum_{k=1}^{K_i}(c_{ik}(t) - d_{ik}(t)) - p_i^\mathrm{u}(t) - \epsilon_i^\mathrm{u}(t) \Big]_+ \\
 + \Big[ p_i^\mathrm{l}(t) + \epsilon_i^\mathrm{l}(t) - p_i(t) - \sum_{k=1}^{K_i}(c_{ik}(t) + d_{ik}(t)) \Big]_+,
 \label{bounds_penalty}
 \end{multline}
where $p_i^\mathrm{u}(t)$ and $p_i^\mathrm{l}(t)$ are the supply bounds during the hour in which $t$ lies, $\epsilon_i^\mathrm{u}(t)$ and $\epsilon_i^\mathrm{l}(t)$ are the sum of the past deviations of the net power injection from the upper and lower bounds during the hour, respectively, which allows the controller to ensure that the average power across the hour can satisfy the bounds. The second term in the objective function for each DER $k$ is
\begin{equation}
 L_{ik}^f(t) = |Q_{ik}(t) - Q_{ik}^{\mathrm{target}}(t)|,
\label{storage_penalty}
\end{equation}
where $Q_{ik}^{\mathrm{target}}(t)$ is the desired state of DER $k$ at time $t$, which depends on the specific objective of the consumer. We assume that the objective is to minimize the cost of electricity based on the given or reduced time-of-use (TOU) rate. Thus, the desired state for each DER is $Q_{ik}^{\mathrm{target}}(t) = Q^{\mathrm{max}}_{ik}(t)$ for $t$ in the 12 hour period before the start of the peak price period (from early morning to late afternoon), and $Q_{ik}^{\mathrm{target}}(t) = Q^{\mathrm{min}}_{ik}(t)$ for $t$ during the peak price period (between late afternoon to late evening). In all other times $t$ (night time) in which the TOU rate is lowest and there is no solar generation, the objective $L_{ik}^f(t) = 0$. 

To determine the limits on the thermal load energy consumption profiles $Q^{\mathrm{min}}_{ik}(t)$ and $Q^{\mathrm{max}}_{ik}(t)$ that correspond to the ends of the desired temperature band, we use the following simple model. We assume that the given baseline thermal load energy consumption profile $Q_{ik}^{\mathrm{base}}(t)$ (see Section~\ref{sec:setup} for how it is determined for the simulations) corresponds to the mid point of the desired temperature band and express the limits on the thermal load energy as
\begin{align*}
Q_{ik}^{\mathrm{max}}(t) &= \min\big\{Q_{ik}^{\mathrm{base}}(t) + \phi\, Q_{ik}^{\mathrm{base}}(T_d), Q_{ik}^{\mathrm{base}}(T_d)\big\},\\
Q_{ik}^{\mathrm{min}}(t) &= \max\big\{0, Q_{ik}^{\mathrm{base}}(t) - \phi\, Q_{ik}^{\mathrm{base}}(T_d), \\
& \quad \quad \quad \quad \quad Q_{ik}^{\mathrm{base}}(T_d) - (T_d - t)\, \delta\, u^{\mathrm{max}}_i \big\}.
\end{align*}
These limits ensure that the total shifted energy consumption does not exceed a fraction $\phi$ of the total thermal energy consumption during the day and is equal to the baseline consumption by the end of the day.

\section{Simulation setup} 
\label{sec:setup}

We simulate and evaluate our DER coordination scheme using the same methodology, data and assumptions detailed in~\cite{joule}. To make the presentation of our results somewhat self-contained, in the following we provide a summary of the simulation setup and the DER bookend controllers.

We use the power flow-optimization system in~\cite{joule} in which the input data includes the linearized network models, load profiles, DER operating parameters, DER penetrations, and prescribed electricity tariffs. Using this data, a network scenario is generated by randomly choosing the EV charging windows and assigning rooftop PV, stationary storage, and flexible loads to the consumer nodes in the network. The operation of the chosen DER control scheme is then simulated to determine the DER power injection profiles for all DERs, which allows us to run quasi-static power flow simulation using OpenDSS~\cite{montenegro2012real} to determine the voltage at each node and apparent power flow through each transformer. This information is then used to compute the transformer and voltage reliability metrics and determine the cost of electricity for each node. This process is repeated for several scenarios and reliability statistics are computed.

As in~\cite{hosting_capacity,joule}, the metric we use for steady-state voltage declares a violation at a node if its voltage magnitude exceeds the specifications in the ANSI C84.1 standard, which represent a deviation of $\pm 5\%$ from the nominal voltage magnitude on average across any 1-hour window. The metric for transformer thermal overloading declares a violation at a transformer if its average apparent power is greater than 120\% of its rated capacity over one or more 2-hour windows.

Table~\ref{tab:network_1} summarizes the main characteristics of the eleven 3-phase distribution networks we used to evaluate our scheme. The columns are the assigned name indicating location and source of network data, type of network data, the number of nodes, transformers, and  consumer nodes for each network. The asterisk by the network name indicates a network that aggregates consumers under the secondary transformer. The networks without asterisk have models for individual consumers under each secondary transformer.

\begin{table}[!h]
\caption{Main characteristics of the networks used in the simulations.}
\label{tab:network_1}
\vspace{3pt}
\centering
\begin{tabular}{llllllllllll}
\hline\\[-6pt]
Name                                                    & Type && $N$&& $N_\mathrm{T}$ && $N_\mathrm{C}$ \\[2pt] \hline\\[-5pt]
Sacramento*~\cite{test_feeders_IEEE}     & Synthetic      && 278                                               && 99                          && 91                           \\[2pt]
Iowa*~\cite{iowa}                          & Real      && 915                                                               && 268                         && 193                           \\[2pt]
Central SF~\cite{grid_models_NREL}       & Synthetic         && 2115                                                             && 232                         && 425                           \\[2pt]
Commercial SF~\cite{grid_models_NREL}    & Synthetic         && 172                                                                && 17                          && 18                           \\[2pt]
Tracy~\cite{grid_models_NREL}            & Synthetic      && 775                                                                   && 108                          && 161                           \\[2pt]
Rural San Benito~\cite{grid_models_NREL} & Synthetic         && 243                                                          && 15                           && 22                            \\[2pt]
Los Banos~\cite{grid_models_NREL}        & Synthetic      && 2010                                                                   && 251                         && 426                           \\[2pt]
Vermont~\cite{EPRI_J1}                    & Real      && 4245                                                     && 828                         && 1384                           \\[2pt]
Arizona*~\cite{test_feeders_IEEE}        & Real     && 138                                                                  && 46                          && 38                           \\[2pt]
Marin~\cite{grid_models_NREL}            & Synthetic      && 3689                                                         && 231                         && 811                           \\[2pt]
Oakland~\cite{grid_models_NREL}       & Synthetic      && 10073                                                        && 658                         && 2426                           \\[2pt] \hline
\end{tabular}
\end{table}

To construct the load profiles, we first use the 2018 residential and commercial consumer load profiles dataset in~\cite{NREL_stock_data} to construct 1-year baseline profiles for each node in each network based on its geographical census code. We then scale these profiles for each node so that its average daily peak load matches that of the network. The reactive power component of the load for each node is chosen such that the power factor is randomly distributed between 0.9 and 0.95 as is commonly assumed, e.g., see~\cite{iowa}.

To determine the electrification increases, we use the projections up to 2050 in~\cite{NREL_EV_adoption}, which is categorized by space heating, water heating, and other loads. We uniformly scale the electrification due to loads other than thermal by their electrification projection percentage. We then randomly choose a consumer that does not already have electric space heating or water heating, and assign to it a load profile of either space or water heating that is converted from the thermal energy in~\cite{NREL_stock_data} to electric energy using the median published efficiency for the corresponding electric appliance in~\cite{CEC_appliance}. This process is repeated until the total added electric energy in a network equals its electrification projection. The sum of air conditioning, space heating, and water heating load profiles at each node represents its baseline thermal load profile referred to in Section~\ref{sec:lc}. We declare a percentage of the thermal load profile as flexible load using the projections in~\cite{flex_load_NREL}, which specifically considers electric furnaces, air conditioners, and water heaters.

EV penetration is defined as the EV charging energy as a percentage of the total energy in the network with no solar included, and assuming penetrations from the high electrification projections in~\cite{NREL_EV_adoption}. We assume the Workplace and home L2 EV chargers to have a rated charging power of 6.3kW~\cite{NREL_EV} and that charging power can be varied between any value less than this rated power. To determine the EV charging windows, we use the synthetic data generator in~\cite{EV_model_siobhan} to produce sample charging schedules each consisting of a starting time, ending time, charging energy demand, and whether charging is done at home or at work over the 1-year simulation horizon. We continue to add new EVs to the network at random until the total energy consumed by EVs as determined by the data generator is equal to the total projected EV electrification penetration energy.

PV penetration is defined as the energy generated by solar PV as a percentage of the total energy consumed by the network, including for EV charging. We set the PV penetration in 2050 for each network to 23\% of the total potential rooftop solar PV generation of its nearest city~\cite{NREL_solar}, which corresponds to approximately the same amount of solar PV capacity as the high penetration scenario for nationwide distributed PV projected in~\cite{NREL_storage}. To determine the PV penetration for the years prior to 2050, we scale the penetrations in proportion to the nationwide PV capacity provided in~\cite{NREL_storage}. The PV generation profiles for each network are obtained from the solar data of the closest geographic region to the network provided in~\cite{NREL_solar_data}. Nodes are assigned a PV generation profile at random until reaching the PV penetration. The profile is scaled such that the ratio of solar PV energy generation to the total energy consumed by the node is randomly distributed between $40-90\%$, which roughly corresponds to the upper quartile estimate of residential PV system sizes given in~\cite{NREL_storage}.

We assume that only nodes with solar PV can have stationary storage and define storage penetration as a percentage of the nodes with PV. The storage penetrations we assume correspond to the highest adoption rates given in~\cite{NREL_storage}, which range between 30\% and 70\% depending on the year. Each storage unit is randomly assigned a capacity which is between $40-80\%$ of the average daily PV energy generation, which corresponds to the upper quartile estimate of residential battery system sizes in~\cite{NREL_storage}, a maximum c-rate of 0.5~\cite{NREL_storage} and a round-trip efficiency of 0.86~\cite{storage_efficiency}. 

The TOU tariffs~\cite{pge_tariffs} used in the simulations include peak, part-peak, and off peak rates for both residential and commercial consumers. The hours during which these rates apply are chosen such that the middle of the peak price hour corresponds to the most frequent total network peak demand hour. In addition, EV TOU rates~\cite{EV_charging_behaviors} are included (see~\cite{joule} for details). We assume that the reduced rate used when the consumers obey their respective supply bounds is 80\% of the regular TOU rates.

\noindent{\bf Bookend controllers}.

To evaluate our bounds scheme, we compare its performance in terms of transformer and voltage violations to two bookend control schemes. The first is a centralized perfect foresight controller that has perfect knowledge of all future loads and DER states 2-days in advance and has full control over all DERs. This centralized controller determines the power injections for all DERs by minimizing a weighted sum of the electricity cost, voltage violations and transformer overloading metrics, and storage operation cost with the grid reliability objectives having the highest weight.
The constraints are the linearized inverse power flow mapping and the storage, EV charging, and flexible load dynamics.

The second bookend control scheme is a local controller that runs fully autonomously at each node and operates its DERs with the goal of minimizing the consumer's total electricity cost. It is a simple heuristic that does not presume knowledge or forecasts of any variable values, except for the end of the current EV charging window and the percentage of thermal load that is flexible, which may be specified by the consumer. At each time step, the controller checks its electricity tariff to find the lowest cost time period and reserves operation to within those time periods as much as possible.

\noindent{\bf Software implementation}.
All data used in the simulations are publicly available and referenced in this paper and in~\cite{joule}. The code for the simulations will be available on Github after publication of this paper.

\section{Simulation results} \label{sec:results}

We simulated our DER coordination scheme using the setup outlined in the previous section for the years 2020 to 2050 using the same scenarios as in~\cite{joule} to provide a fair comparison of our scheme to the two bookend controllers. 

\begin{figure}[htpb]
\centering
\includegraphics[scale=0.7]{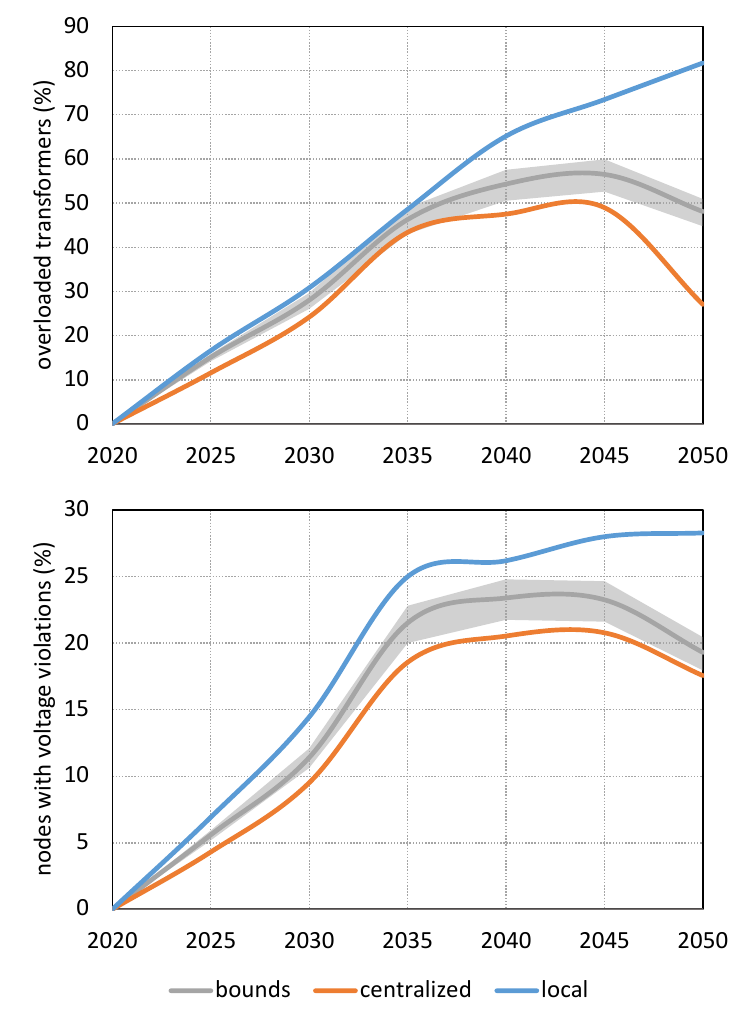}
\caption{Aggregated means and standard errors of the violation percentages with the bounds scheme compared to the means of the bookend control schemes for each simulation year.}
\label{fig:years}
\end{figure}

Figure~\ref{fig:years} plots the means and standard errors across the 11 networks for the transformer overloading and voltage violation metrics between 2020 and 2050 using our bounds scheme as well as the means with the bookend controllers. Note that from 2020 to 2030, the transformer overloading and voltage violations steadily increase for all three control schemes. From 2035 to 2045, the bounds scheme and centralized controller see a slower increase in violations as the amount of flexibility and stationary storage begin to offset the downsides of increased electrification. Finally, after 2045, the amount of load flexibility becomes high enough to actually decrease the violations for both the bounds scheme and the centralized controller. In 2050, the means of the transformer and voltage violations for the bounds scheme are 49\% and 20\%, respectively, compared to 29\% and 18\% for the centralized controller and 81\% and 28\% for the local controller. Hence the bounds scheme captures 62\% and 75\% of the violation reductions of the centralized controller versus the local controller.

\begin{figure}[htpb]
\centering
\includegraphics[scale=0.69]{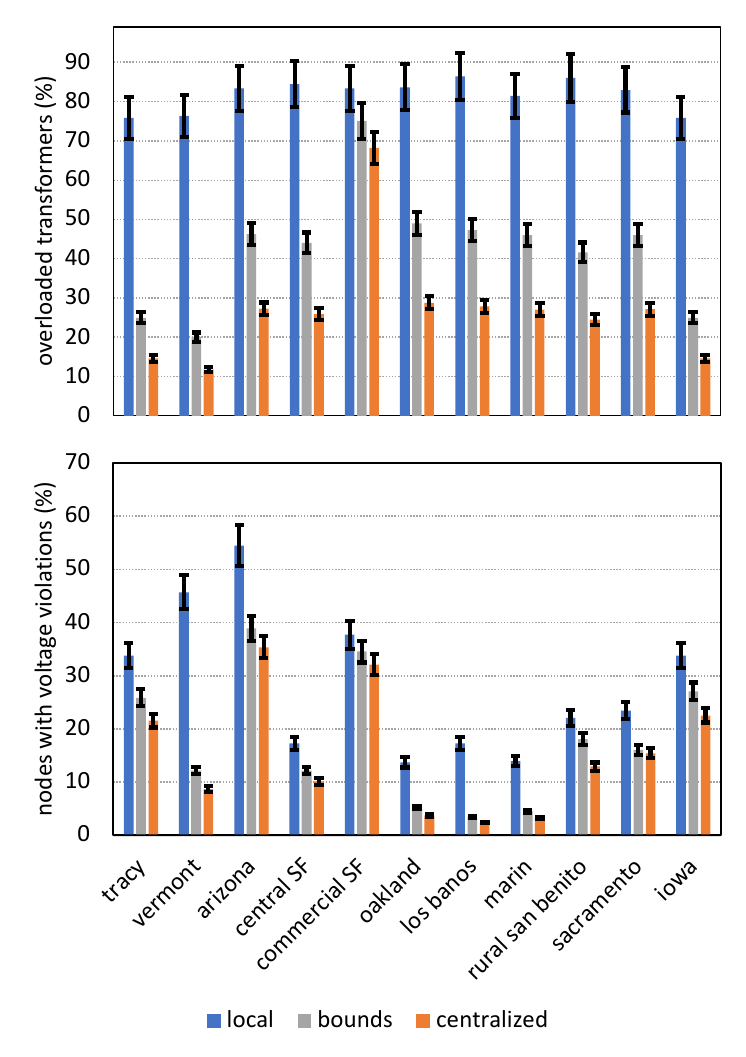}
\caption{Means and standard errors of the violation percentages for individual networks in 2050.}
\label{fig:bars}
\end{figure}

Figure~\ref{fig:bars} shows the means and standard errors of the violation percentages for each individual network averaged across all 16 scenarios in 2050. The bounds scheme is able to reduce the number of overloaded transformers significantly across all networks with the exception of the commercial SF network, which has much less flexible load due to having only commercial consumers as opposed to primarily residential consumers as for all other networks. Note that all networks exhibit noticeable reduction in voltage violations with the bounds scheme. However, there are significant variations across networks because some are naturally more prone to voltage violations than others due to their specific configuration of power lines and consumers.

\begin{figure}[htpb]
\centering
\includegraphics[scale=0.5]{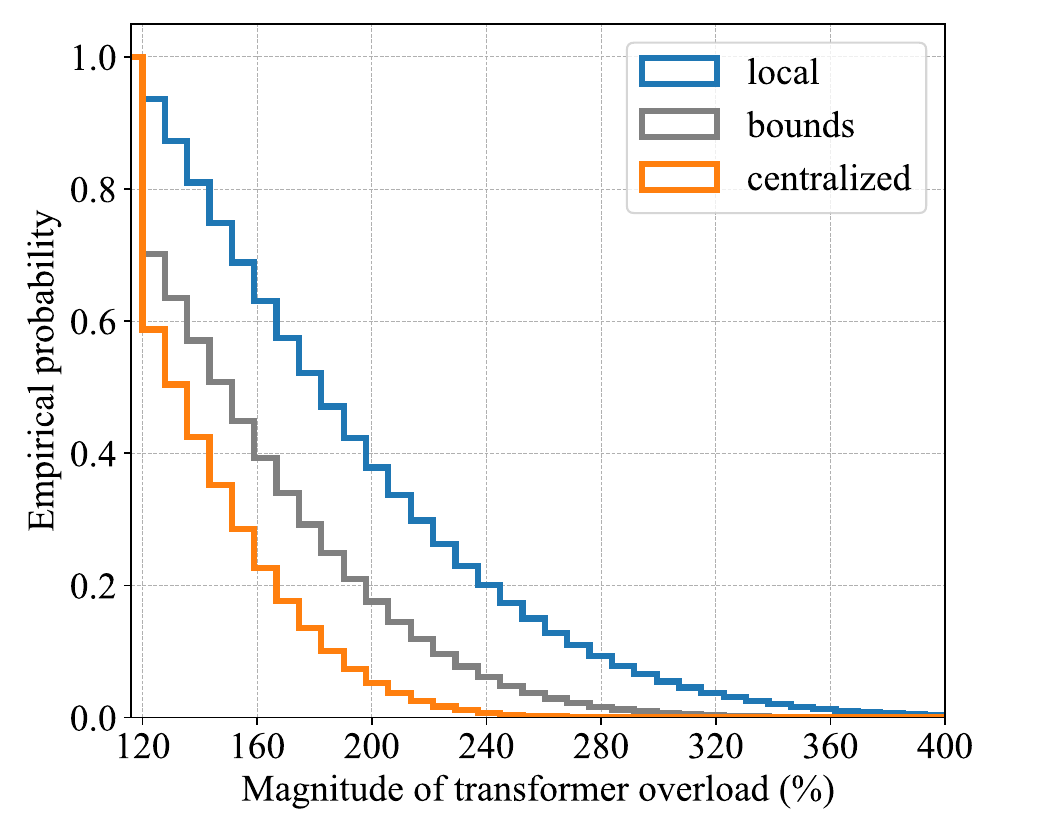}
\caption{The empirical probability of the percentage magnitude of transformer apparent powers being greater than value on the x-axis across all networks and scenarios in 2050.}
\label{fig:hist}
\end{figure}

As pointed out in~\cite{joule}, DER coordination provides benefits to reliability beyond reducing the transformer and voltage violations according to the binary-valued metrics. Figure~\ref{fig:hist} plots the empirical probability that the percentage magnitude of transformer apparent powers across all networks and scenarios in 2050 being greater than value on the x-axis. Note that shifting the distribution of the magnitudes of all transformer apparent powers towards the y-axis represents less violations. The bounds controller has an average magnitude of 192\% for transformer overloading events as opposed to 243\% to 159\% for the fully local and fully centralized controllers. The results for voltage violations are similar with the bounds controller having an average magnitude of 5.47\% as opposed to 5.93\% to 5.32\% for the local and centralized controllers. 

\subsection{Impact on peak load}
\begin{figure}[htpb]
\centering
\includegraphics[scale=0.7]{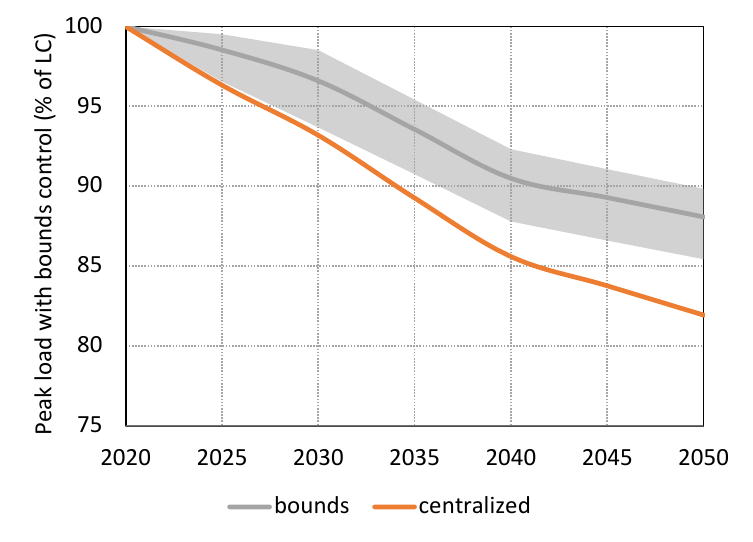}
\caption{Plot of the means and standard errors of the peak load over the simulation horizon as a percentage of the bookend local controller peak load.}
\label{fig:peak_load}
\end{figure}

In addition to reducing the number of transformers that experience overloading and nodes with voltage violations, the bounds scheme reduces the peak network load over the simulation horizon relative to the local controller. As shown in Figure~\ref{fig:peak_load}, the peak load with the bounds scheme is reduced by 12\% on average in 2050 compared to the bookend local controller versus an 18\% decrease from the centralized control scheme.

\subsection{Impact on electricity cost}
In~\cite{joule}, we showed the centralized controller increases electricity cost by 5.1\% in 2050 over the local controller, which focuses only on minimizing electricity cost. With the bounds controller, we find that when calculating the cost with no incentive to follow the bounds, the increase is between 9.8\% and 11.1\% relative to the local controller. When factoring in the 80\% incentive we assumed in our simulations, the cost of electricity is reduced to between 93.4\% and 95.0\% of the cost under the local bookend controller.


\subsection{Impact of load flexibility}

\begin{figure}[htpb]
\centering
\includegraphics[scale=0.69]{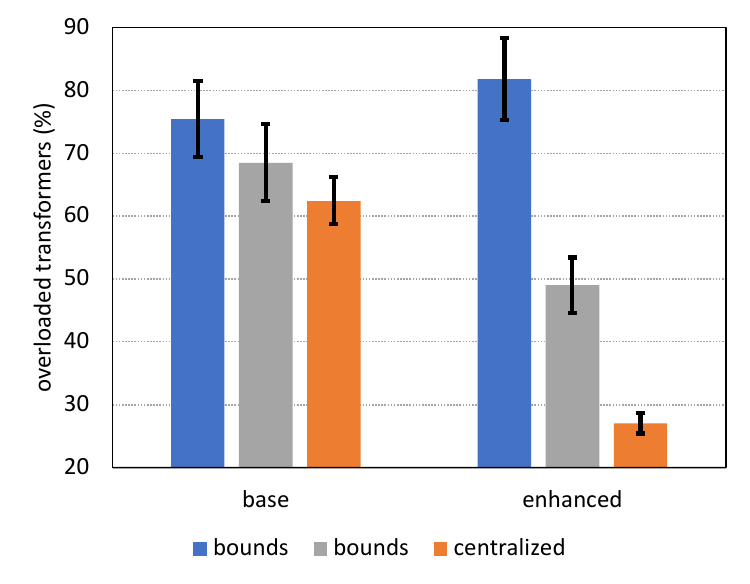}
\caption{Means and standard errors of the percentge of overloaded transformers across all networks in 2050 with base and enhanced load flexibility.}
\label{fig:flex}
\end{figure}

The projections of load flexibility from~\cite{flex_load_NREL} include a base case and an enhanced case with the enhanced one, which we assumed in the above results, projecting significantly more load flexibility. In order to determine the impact of the amount of load flexibility on the performance of the bounds scheme, Figure~\ref{fig:flex} shows the mean and standard errors of the percentage of overloaded transformers averaged across all networks and scenarios in 2050 with the three control schemes for the base and enhanced flexibility cases. The means of the percentage of overloaded transformers for the bounds scheme and the base load flexibility are 69\% and 25\%, respectively, compared to 49\% and 20\% with enhanced load flexibility. Thus, load flexibility represents a significant contribution to the performance of the bounds scheme.

\section{Conclusion}

We presented a DER coordination scheme which aims to protect distribution grids with high penetrations of DER and electrification. We demonstrated that this scheme can capture a significant fraction of the reliability benefits of a perfect foresight centralized controller relative to current autonomous DER control~\cite{joule}. Another important benefit of our scheme relative to autonomous DER control is that it reduces peak network power even though this objective is not explicitly included in the bounds problem formulation. We showed our scheme addresses all the requirements of an implementable DER coordination.

Our bounds scheme can be readily deployed using existing computing and communication infrastructure and does not require the development or deployment of any new hardware (the GC and LC operations can both be handled using existing cloud infrastructure and DER embedded controllers) or grid infrastructure, making it very cost effective especially when considering the significant savings it provides to grid infrastructure upgrades. Deploying our scheme, however, requires the involvement of distribution grid operators (DSO) not only to provide the necessary information about the grid, e.g., the linearized grid models used in this paper, but also to ensure broad consumer adoption of the scheme through various incentives, such as our proposed reduced TOU rate. Our scheme can also operate seamlessly with existing demand response and virtual power plants (VPP) programs which aim to reduce electricity cost during extreme climate conditions. In such cases, the LCs would adjust their control objectives to satisfy the requests for grid services from the DSO, DER vendors or third party DER aggregators while still aiming to keep power injections within the supply bounds.  

To evaluate the bounds scheme, we assumed that the demand bounds are determined by the GC using historical smart meter data. Other ways to determine these bounds include incentivizing consumers to submit their own demand bound forecasts. Having more accurate demand bounds would allow the GC to better allocate supply bounds to other nodes who may need the additional capacity. We also assumed that all LCs use the same control objective. The scheme, however, allows consumers to have different objectives provided they include staying within the given supply bounds. 

The 80\% of TOU incentive assumed in this paper is only meant to illustrate the potential for our scheme to provide savings to both the consumers who use their DER flexibility for grid reliability and the DSO who saves on distribution asset upgrade costs. The exact values of the incentive will have to be calculated on an individual network basis by considering the cost savings of deferring transformer and voltage upgrades and differences in cost of electricity over time. These costs can vary widely depending on the DSO network equipment, location, and electricity generation portfolio.

\section*{Acknowledgment}
The authors would like to thank Anthony Degleris for bringing to their attention the maximum volume axis aligned box problem in~\cite{boyd_vandenberghe_2004}. The work presented in this paper was supported under ARPA-E award DE-AR0000697 and US DOE award DE-OE0000919.

\bibliographystyle{IEEEtran}
\bibliography{refs.bib}

\end{document}